\documentclass[aps,pra,showpacs, twocolumn]{revtex4}

\usepackage{graphicx}
\usepackage{amssymb}
\usepackage{amsmath}
\usepackage{colordvi}
\usepackage{color}

\newcommand{\PT}{{\cal PT}}

\newcommand{\gcr}{\gamma_{cr}^{(1)}}
\newcommand{\ggcr}{\gamma_{cr}^{(2)}}

\newcommand{\bw}{{\bf w}}
\newcommand{\bW}{{\bf W}}
\newcommand{\tbw}{\tilde{\bf w}}
\newcommand{\br}{{\bf r}}

\newcommand{\tb}{\tilde{b}}

\newcommand{\T}{{\cal T}}
\newcommand{\p}{{\cal P}}

\newcommand{\diag}{\mbox{diag}}
\newcommand{\vep}{\varepsilon}
\begin{document}

\title{ Parity-time symmetric optical coupler with birefringent arms}

\author{K. Li$^*$, D. A. Zezyulin$^{\dag}$, V. V. Konotop$^{\dag}$, P. G. Kevrekidis$^*$}

\affiliation{$^*$ Department of Mathematics and Statistics, University of Massachusetts,
Amherst, Massachusetts 01003-4515, USA\\
 $^\dag$Centro de F\'{\i}sica Te\'orica e Computacional, and Departamento de F\'{\i}sica, Faculdade de Ci\^encias,
 Universidade de Lisboa,
Avenida Professor Gama
Pinto 2, Lisboa 1649-003, Portugal
}

\date{\today}

\begin{abstract}
 In this work, we propose a parity-time ($\PT$-) symmetric   optical coupler whose arms are birefringent
waveguides as a realistic physical model which leads to a
so-called quadrimer i.e., a four complex field setting.
We seek stationary solutions of the resulting
linear and nonlinear model, identifying its linear point of $\PT$
symmetry breaking and examining the corresponding nonlinear
solutions  that persist up to this point, as well as, so-called,
ghost states that bifurcate from them. We obtain the relevant
symmetry breaking bifurcations between symmetric (circularly polarized) 
and asymmetric (elliptically polarized) states and numerically follow the
associated dynamics which give rise to growth/decay even within
the $\PT$-symmetric phase. Our symmetric
stationary nonlinear
solutions are found to terminate in saddle-center bifurcations
which are analogous to the linear $\PT$-phase transition.
We found that the $\PT$ symmetry significantly changes the stability and dynamical properties  of the modes with different polarizations.
\end{abstract}


\maketitle

\section{Introduction}
\label{intro}

An optical coupler with one arm having losses and another one having gain, balanced against each other, recently became a test-bed for many phenomena
originating from the interplay of the  parity-time ($\PT$) symmetry and nonlinearity.
Unidirectional dynamics~\cite{kot1}, unversality of the dynamics~\cite{kot2},
symmetry breaking properties~\cite{kip,pgk},
switching of the beams~\cite{sukh1} and of solitons~\cite{R30add3}, formation of symmetric and asymmetric bright
solitary waves~\cite{R30add1,R30add2},  breathers~\cite{baras},
and their stability~\cite{R30add4}, dark solitons~\cite{BKM}, as well as the emergence of
ghost states~\cite{R46a,R46b,R46}
and large-scale temporal  $\PT$-symmetric lattices~\cite{christo_nat2}
are some among the many topics that have been
touched upon in very intense recent theoretical and experimental
work.

As a direct extension of the previous activity, a large chunk of
which has focused on the prototypical setting of the
$\PT$-symmetric dimer, there emerges a problem  of   effect of
$\PT$ symmetry and nonlinearity on the polarization of the
electric field. In that regard, the previously proposed settings, to the best of
our knowledge, were chiefly focused on  effectively scalar models. On the
other hand, the ``vector'' type of problems is natural for
experimental settings where the exploited fibers obey
birefringence, since the two orthogonal polarizations are to be
taken into account~\cite{Menyuk}.

In this work we consider a $\PT$-symmetric coupler whose arms are
birefringent waveguides.  Assuming that the first waveguide is
active and the second one is absorbing,  we address the problem of
a $\PT$-preserving (in the linear limit) configuration. While being an
interesting model from a physical point of view, this setting also
offers a different (in comparison to what was studied before)
mathematical situation where the nonlinear modes bifurcate from
doubly degenerated eigenvalues of the linear problem. This
requires the generalization of earlier developed approaches (e.g.
like the one reported in \cite{ZK}) for the bifurcation of the
nonlinear modes from the linear spectrum. In addition, it presents
a rich playground for dynamical systems analysis, due to the
emergence of a variety of saddle-center bifurcations (nonlinear
analogs of the linear
 $\PT$-phase transition), as well as symmetry-breaking (pitchfork) ones.
It is these nonlinear states, their emergence, stability, dynamics
and the asymptotics of the system that we will focus on hereafter.

The organization of the paper is as follows. In section~\ref{sec:model}, we present
the model in its evolution as well as in its stationary form.
In section~\ref{sec:lin}, we focus on its linear properties. Then, in section~\ref{sec:nonlin},
we examine the nonlinear modes (and bifurcations). Finally, in
section~\ref{sec:stab}, we briefly touch upon the dynamical implications of our
findings and in section~\ref{sec:Concl}, we present some conclusions, as well
as some potential directions for future work.

\section{The model}
\label{sec:model}

\begin{figure}[h]
\centerline{
       \includegraphics[width=0.45\columnwidth]{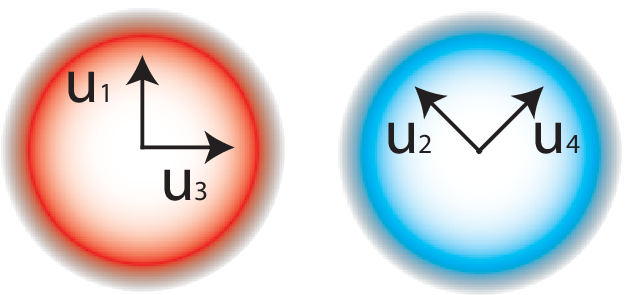}
       \hspace{1.1cm}
       \includegraphics[width=0.35\columnwidth]{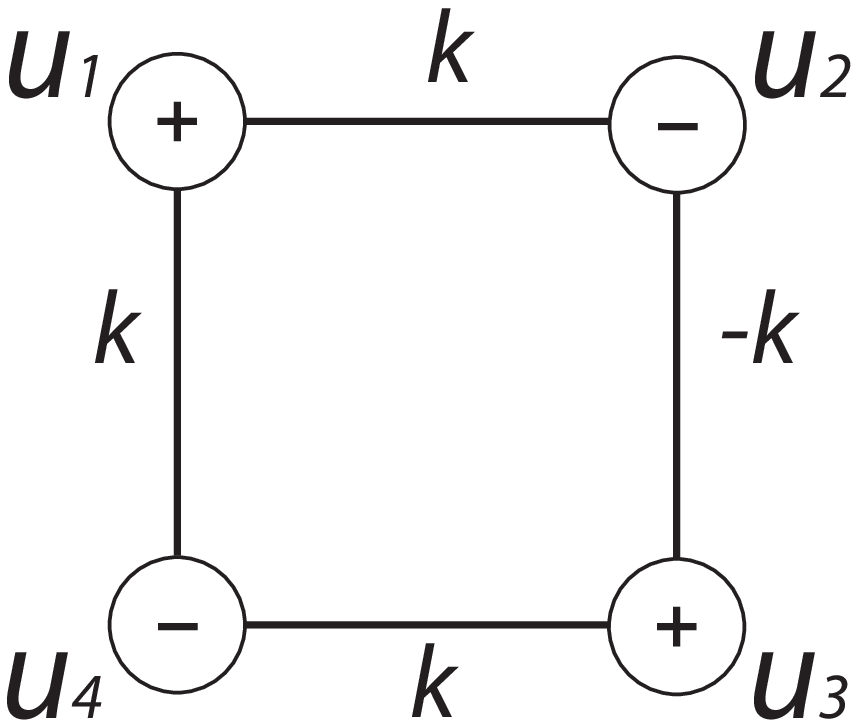}}
\caption{(Color online) (a) Schematic presentation of a $\PT$-symmetric coupler based on birefringent fibers.   (b) Equivalent graph (plaquette) representation
illustrating the $\PT$-symmetry. Here $-$ and $+$ stand for
lossy and active waveguides, respectively.} \label{fig:coupler}
\end{figure}

We specify the problem by imposing that the principal optical axes
of the two Kerr-type waveguides are $\pi/4$-rotated with respect to each other, as it is schematically
represented in Fig.~\ref{fig:coupler}. In each arm, labeled by $j$,  there are two orthogonal
field components of the electric fields which we write down in the form~\cite{Menyuk} ($j=1,2$):
\begin{eqnarray}
{\bf E}_j({\bf r}, z, t)=
 \left[u_j(z)A_j(\br-\br_j)e^{-i\beta_{j}z}{\bf e}_{j}
\right. \nonumber \\
+
\left. u_{j+2}(z)A_{j+2}(\br-\br_j)e^{-i\beta_{j+2}z}{\bf e}_{j+2}\right]\sqrt{\frac{2}{\chi}}e^{i\omega t}+c.c.
\end{eqnarray}
Here $u_j$ are the field envelopes depending on the propagation distance $z$,
i.e. we consider the stationary -- in time -- problem, assuming that the   carrier wavelength $\lambda_0$ is in the region of the normal group velocity dispersion, thus ruling out a possibility of modulational instability; $\br=(x,y)$ is a transverse radius vector, and $\br_{1,2}$ are the positions of the centers of the cores of the coupler. The real parameters $\beta_j$ are the propagation constants of each of the field components, and ${\bf e}_{j}$ are the polarization vectors, which are mutually orthogonal in each arm of the coupler, i.e. ${\bf e}_{1}\cdot {\bf e}_{3}={\bf e}_{2}\cdot {\bf e}_{4}=0$.
The real functions $A_j(\br-\br_j)$ and $A_{j+2}(\br-\br_j)$ describe the transverse distributions of the fields in each waveguide and the normalization coefficient $\sqrt{2/\chi}$, where $\chi$ is the Kerr coefficient, is introduced for convenience. For $j=1$ and $j=2$ the functions $A_j(\br)$ are centered in different points $\br_j$. Also, for the sake of simplicity, we consider 
$A_j(\br)=A_{j+2}(\br)= A(\br)$ (for $j=1,2$), such that the integral
$$
({\bf e}_{j}\cdot {\bf e}_{j+1}) \frac{\int A(\br-\br_j)A(\br-\br_{j+1})d^2{\bf r}}{\int A^2(\br)d^2{\bf r}}
$$
(the  integration is performed over the transverse plane)
describes the linear coupling between the respective modes.  Since in the configuration shown in Fig.~\ref{fig:coupler} ${\bf e}_{1}\cdot {\bf e}_{2}={\bf e}_{1}\cdot {\bf e}_{4}={\bf e}_{3}\cdot {\bf e}_{4}=-{\bf e}_{3}\cdot {\bf e}_{2}=1/\sqrt{2}$ we  use the single linear coupling coefficient $k$ (see also~\cite{coupling}).
Then following the analysis  described in details in~\cite{Menyuk} we end up with the system of equations:
\begin{subequations}
\label{dynam}
\begin{eqnarray}
i\frac{du_1}{dz}=-k(u_2+u_4)+i\gamma u_1-\left(|u_1|^2+\frac 23 |u_3|^2\right)u_1
\nonumber
\\
-\frac 13 u_3^2u_1^*e^{i\Delta_1z}
\\
i\frac{du_2}{dz}=-k(u_1-u_3)-i\gamma u_2-\left(|u_2|^2+\frac 23  |u_4|^2\right)u_2
\nonumber
\\
 - \frac 13 u_4^2u_2^*e^{i\Delta_2z}
\\
i\frac{du_3}{dz}=-k(u_4-u_2)+i\gamma u_3-\left(\frac 23  |u_1|^2+|u_3|^2\right)u_3
\nonumber
\\
  -\frac 13 u_1^2u_3^*e^{ -i\Delta_1z}
\\
i\frac{du_4}{dz}=-k(u_1+u_3)-i\gamma u_4-\left(\frac 23  |u_2|^2+|u_4|^2\right)u_4
\nonumber
\\
  - \frac 13 u_2^2u_4^*e^{-i\Delta_2z}
\end{eqnarray}
\end{subequations}
Here $\gamma>0$ describes gain in the first waveguide and dissipation in the second waveguide,
$\Delta_j=\frac{4\pi c}{\lambda_0}\left(\beta_j^\prime-\beta_{j+2}^\prime\right)$ with $\beta_j^\prime=\frac{d\beta_j(\omega_0)}{d\omega_0}$, $\omega_0$ being the carrier wave frequency, is a properly normalized mismatch between the propagation constants of the orthogonal polarizations $u_{j+2}$ and $u_j$. The
asterisk stands for complex conjugation.

We will be interested in the stationary solutions, in particular in their linear
stability properties and ensuing nonlinear dynamics which can be found in the
two prototypical limiting cases  of (i)  zero mismatches $|\beta_{j}-\beta_{j+2}|=0$
and (ii) large mismatches  $|\beta_{j}^\prime-\beta_{j+2}^\prime|\gg  k\lambda_0/c$ when the respective nonlinear terms can be neglected. It is convenient to introduce a parameter $\alpha$
which vanishes ($\alpha=0$) in the case (ii) and is unity ($\alpha=1$) in the
case (i). Using the standing wave ansatz $u_j(z) =  w_je^{ibz}$, where
$w_j$ are $z-$independent, we obtain the system of algebraic equations:
\begin{subequations}
\label{static}
\begin{eqnarray}
    bw_1 =  k(w_2+w_4) -i\gamma w_1 + \left(|w_1|^2+\frac 23|w_3|^2\right)w_1
    \nonumber \\ + \frac \alpha 3   w_3^2w_1^*,
    \\
    bw_2 = k(w_1-w_3) +i\gamma w_2 + \left(|w_2|^2+ \frac 23 |w_4|^2\right)w_2
    \nonumber \\ + \frac \alpha 3 w_4^2w_2^*,
    \\
    bw_3 = k(w_4-w_2) -i\gamma w_3 + \left(\frac 23|w_1|^2+|w_3|^2\right)w_3
    \nonumber \\ + \frac \alpha 3 w_1^2w_3^*,
    \\
    bw_4 = k(w_1+w_3)+i\gamma w_4  + \left(\frac 23|w_2|^2+|w_4|^2\right)w_4
    \nonumber \\+ \frac \alpha 3 w_2^2w_4^*.
\end{eqnarray}
\end{subequations}
Below the spectral parameter $b$ will be also referred to as the propagation constant.


\section{Properties of the linear problem}
\label{sec:lin}%

First we address the underlying  linear problem [which corresponds to the situation when  all cubic terms in Eqs.~(\ref{static}) are negligible]. It can be rewritten in the matrix form $\tilde{b} \tilde{{\bf w}} ={H} \tilde{{\bf w}}$ where
\begin{eqnarray*}
\tilde{{\bf w}} =\left(\!\!
         \begin{array}{c}
           \tilde{w}_1  \\ \tilde{w}_2  \\ \tilde{w}_3 \\ \tilde{w}_4
         \end{array}
       \!\! \right)
\,\,\,\mbox{and}\,\,\,
{ H}=\left(\!\!
         \begin{array}{cccc}
           -i\gamma & k & 0 & k\\
           k & i\gamma & -k & 0\\
           0 & -k & -i\gamma & k \\
           k & 0 & k & i\gamma
         \end{array}
       \!\right)
\end{eqnarray*}
(hereafter we use tilde in order to distinguish eigenvalues and eigenvectors of the linear problem).

The operator $H$ is $\PT$ symmetric, which means that $[H, \PT] = H\PT - \PT H=0$, where $\p$ is a spatial reversal linear operator
\begin{eqnarray}
\label{eq:P}
{\p}=\left(\!\!
         \begin{array}{cccc}
           0 & 0 & 0 & 1\\
           0 & 0 & 1 & 0\\
           0 & 1 & 0 & 0 \\
           1 & 0 & 0 & 0
         \end{array}
       \!\right)
\end{eqnarray}
and $\T$ performs element-wise complex conjugation: $\T \bw = \bw^*$. The spectrum of operator $H$ consists of \textit{two double eigenvalues}
\begin{eqnarray}
\label{b}
\tb_\pm=\pm\sqrt{2k^2-\gamma^2},
\end{eqnarray}
  which are real for $\gamma<\gamma_{cr}^{(1)}$ where $\gcr=\sqrt{2}k$ will be referred to as a primary critical point: the spontaneous $\PT$ symmetry breaking occurs at $\gamma_{cr}^{(1)}$ above which the eigenvalues are all imaginary.
In order to visualize the $\PT$ symmetry of the linear system, following~\cite{ZK} one can represent it with a graph shown in the right panel of Fig.~\ref{fig:coupler}, reminiscent of four linearly coupled waveguides~\cite{ZK} (notice however the sign difference in the coupling constants) or plaquettes~\cite{guenther}.

Since the details of our analysis are  the same for both eigenvalues $\tb_-$ and $\tb_+$, we drop the subscripts $+$ and $-$ wherever this does not lead to confusion. Despite having double eigenvalues in its spectrum, $H$ is diagonalizable below the $\PT$-symmetry breaking point. This means that double eigenvalues are semisimple,  i.e.    for an eigenvalue  $\tb$  one can find  two linearly independent eigenvectors, i.e.  $H\tbw^{(j)} = \tb\tbw^{(j)}$, where $j=1,2$. Moreover,  each eigenvalue   $\tb$   possesses an invariant subspace spanned by  $\tbw^{(1)}$  and $\tbw^{(2)}$.

Let us also  notice  the following peculiarity of the case at hand. In a situation where a $\PT$-symmetric operator $H$ has no multiple eigenvalues, the condition of unbroken $\PT$ symmetry (i.e. reality of all the eigenvalues) requires that for each eigenvalue $\tb$ the corresponding eigenvector $\tbw$ can be chosen as an eigenstate of the $\PT$ operator, i.e. $\PT\tbw = \tbw$. However, in the situation at hand
%
arbitrarily chosen linearly independent eigenvectors $\tbw^{(1)}$ and $\tbw^{(2)}$ may not be  $\PT$ eigenstates. However, unbroken $\PT$ symmetry requires that a certain \textit{linear combination} of $\tbw^{(1)}$ and $\tbw^{(2)}$  is an eigenstate for the $\PT$ operator. More specifically,  it is easy to establish that   all the eigenvectors  that belong to the invariant subspace of $\tb$ and, at the same time,   are the  eigenstates for the $\PT$ operator, can be parametrized by a complex parameter $a$ as follows:
\begin{equation}
\label{eq:PTeigs}
\tbw = \left(\!\!
         \begin{array}{c}
           a^*\\
           \frac{ia^*(\gamma-i\tb)}{k} - a\\
           \frac{-ia(\gamma+i\tb)}{k} - a^*\\
           a
         \end{array}
       \!\right)
\end{equation}
Being interested in linearly independent vectors $\tbw$,  it is sufficient to consider only the vectors with
\begin{eqnarray}
a=e^{i\theta}.
\label{a}
\end{eqnarray}
 Then Eq.~(\ref{eq:PTeigs}) yields a monoparametric set  of eigenvectors $\tbw(\theta)$ with a real parameter $\theta$.  In particular, setting $\theta^{(1)}=0$ and $\theta^{(2)}=\arctan(\frac{2k-\tb}{\gamma})$ one can choose two orthogonal (and therefore linearly independent)  eigenvectors:
\begin{equation}
\tbw^{(1,2)} = \tbw(\theta^{(1,2)}), \quad \langle \tbw^{(1)}, \tbw^{(2)}\rangle = 0
\end{equation}
 (hereafter we use the standard scalar product $\langle {\bf g}, {\bf h}\rangle = \sum_{j=1}^4 g_j h_j^*$).
Any eigenvector $\tbw$ corresponding to the eigenvalue $\tb$ can be represented as a linear combination of  $\PT$ eigenstates $\tbw^{(1,2)}$: $\tbw = \lambda_1\tbw^{(1)} + \lambda_2\tbw^{(2)} $ (of course, this does not mean that any eigenvector $\tbw$ is also a $\PT$ eigenstate).

Let us also introduce a Hermitian adjoint operator $H^\dag$.
Since the matrix $H$ is symmetric,  one has $H^\dag=H^*$. As long as
$\PT$ symmetry of $H$ is unbroken, the spectrum of the adjoint
operator  $H^\dag$ also consists of two double eigenvalues $b_\pm$
which are semi-simple. Any eigenvector corresponding to an
eigenvalue $\tb$ of the adjoint operator $H^\dag$ can be
represented as a linear combination of $(\tbw^{(1)})^*$ and
$(\tbw^{(2)})^*$.

\section{Nonlinear modes}
\label{sec:nonlin}
\subsection{Bifurcations from the linear limit}
\label{subsec:nonlin}
Now we
develop a perturbation
theory for the eigenstates of the linear problem giving rise to
monoparametric families of nonlinear modes. We will look for nonlinear modes $\bw$ that are eigenstates of the $\PT$ operator, i.e. $\PT \bw = \bw$.  To this end
we introduce the expansions
\begin{equation}
\label{expansion}
\bw =
\vep\tbw(\theta) + \vep^3 {\bW}_3+ \ldots\quad\mbox{and}\quad  b = \tb + \vep^2
B_2 + \ldots
\end{equation}
Here $\vep$ is a small  real parameter, $\bW_3$ and $B_2$
are the coefficients of the expansions and $\theta$ is to be
determined from the symmetry of the solution (see below). We notice that for the expansion to be meaningful the coefficient $B_2$ must be real.

Expansions (\ref{expansion}) describe nonlinear modes that bifurcate  from the \textit{linear
limit}  corresponds to $\vep=0$ and is given by the eigenvector
$\tbw(\theta)$ being a linear combination of $\tbw^{(1,2)}$ such that $\PT\tbw(\theta)=\tbw(\theta)$. Respectively, in the linear limit the parameter $b$ is given by
the eigenvalue $\tb$.

Passing from $\vep=0$ to $0<\vep\ll 1$ one has to compute the
coefficients $\bW_3$ and $B_2$. While the physical sense of the coefficient $B_2$ is clear --- it is a deviation of the propagation constant due to small nonlinearity, it turns out that $B_2$ also has a clear geometrical interpretation. Indeed, let us  consider the total energy flow through the coupler, which is defined by
\begin{equation}
U=\sum_{j=1}^4 |w_j|^2,
\end{equation}
Expansions (\ref{expansion}) imply that in the vicinity of the linear limit $U = \vep^2\langle  \tbw(\theta), \tbw(\theta) \rangle + O(\vep^3)$. Therefore,   the coefficient $B_2$ governs a slope of the energy curve in the vicinity of the bifurcation point, i.e.
\begin{equation}
\label{eq:slope}
\left. \frac{\partial U}{\partial b}\right|_{b=\tb, U=0} = \frac{\langle  \tbw(\theta), \tbw(\theta) \rangle}{B_2}.
\end{equation}

For the sake of definiteness, now we concentrate on the case $\alpha=0$. Then the nonlinear problem (\ref{static}) is conveniently written in the matrix form
\begin{equation}
\label{eq:matrixfoem}
b\bw = H \bw + F(\bw) \bw,
\end{equation}
where $F(\bw)$ is a diagonal matrix-function describing the nonlinearity:
$F(\bw) = \diag (|w_1|^2+\frac 23|w_3|^2, |w_2|^2+\frac 23|w_4|^2, \frac 23|w_1|^2 + |w_3|^2, \frac 23|w_2|^2+|w_4|^2)$. Substituting (\ref{expansion})  into Eq.~(\ref{eq:matrixfoem}), noticing that  $F(\bw) = \vep^2F(\tbw)+O(\vep^3)$,  and collecting the terms order of $\vep^3$, we obtain
\begin{equation}
\label{eq:vep3}
(H-\tb) \bW_3 = -[F(\tbw(\theta)) - B_2]\tbw(\theta).
\end{equation}
Equation (\ref{eq:vep3}) implies two possibilities.  The first one
corresponds to the case when  at some $\theta$   the eigenvector
$\tbw(\theta)$  of the operator $H$ is simultaneously  an
eigenvector for  the matrix $F(\tbw(\theta))$.   Then the
coefficient $B_2$ can be chosen  as an eigenvalue of
$F(\tbw(\theta))$ corresponding to the eigenvector $\tbw(\theta)$
(provided that this  eigenvalue is real).   In this situation, the right
hand side
of Eq.~(\ref{eq:vep3}) is zero and it is sufficient to set $\bW_3
= 0$. Since the matrix $F(\tbw(\theta))$ is diagonal, its
eigenvalues are equal to its diagonal elements and the
corresponding eigenvectors are given as columns of the $4\times 4$
identity matrix.

Let us first assume that all the eigenvalues of
$F(\tbw(\theta))$ are simple. In this case $\tbw(\theta)$ can not
be an eigenvector for  $F(\tbw(\theta))$. The latter fact becomes
evident if one notices that $\tbw(\theta)$ has no zero entries for
any $\theta$ [see the definition (\ref{eq:PTeigs})]. Therefore, 
$\tbw(\theta)$ can be an eigenvector
for $F(\tbw(\theta))$ only if $F(\tbw(\theta))$ has a multiple
eigenvalue.  Then $\tilde{\textbf{w}}(\theta)$ could be searched 
in the form of a linear combination of eigenvectors corresponding 
to the multiple eigenvalue. However, using the same argument, i.e.
the fact that all entries of the vector $\tbw(\theta)$ are
nonzero,  one can see that even if $F(\tbw(\theta))$ has a double
or an triple eigenvalue, the matrix $F(\tbw(\theta))$ still can
not have $\tbw(\theta)$ among its eigenvectors.  Therefore
$\tbw(\theta)$ can be an eigenvector of $F(\tbw(\theta))$ only if
all its eigenvalues are equal. Imposing this constraint on
the matrix $F(\tbw(\theta))$, one obtains $|w_1| = |w_3|$ and
$|w_2| = |w_4|$. Noticing that the 
form of $\tilde{\textbf{w}}(\theta)$ implies, through Eqs.~(\ref{eq:PTeigs}) and (\ref{a}), that
$|w_1| = |w_4| = 1$ we conclude  that $\tbw(\theta)$ is an
eigenvector of $F(\tbw(\theta))$ only if the moduli of  all the entries
of $\tbw(\theta)$  are equal to unity. Then matrix
$F(\tbw(\theta))$ is equal to the  $4\times 4$ identity matrix
multiplied by $5/3$.  Requiring the  moduli of  all the entries
of $\tbw(\theta)$  to be equal, one arrives at the equation for
$\theta$ whose root is given as
\begin{equation}
\label{eq:thetasymm}
\theta =  \frac \pi8 - \frac{1}{2}\arctan\left(\frac \tb\gamma\right).
\end{equation}
Since now  the moduli of all the entries of $\tbw(\theta)$ are equal to
unity and therefore  $\langle  \tbw(\theta), \tbw(\theta) \rangle
=4$, $B_2=5/3$ and Eq.~(\ref{eq:slope}) readily yields  that  in the
vicinity of $\tb$  the slope    $\left. \partial U /\partial b
\right|_{b=\tb, U=0} = 12/5=2.4$. Notice that the found value does
not depend on $k$ or $\gamma$, and thus
these modes correspond to their counterpart in pure conservative coupler
with birefringent arms with $\gamma=0$ [Eq.~(\ref{eq:thetasymm}) is   valid in this case since $\arctan(\pm\infty)=\pm\pi/2$].

Let us now consider the second possibility to fulfill
Eq.~(\ref{eq:vep3}).  If for  some  $\theta$ the corresponding
$\tbw(\theta)$ is not  an eigenvector for  $F(\tbw(\theta))$, then
one must satisfy Eq.~(\ref{eq:vep3}) choosing nonzero $\bW_3$.
Then the coefficient $B_2$ is to be determined from the
solvability condition which requires   the right hand side of
Eq.~(\ref{eq:vep3})  to be  orthogonal to all the  eigenvectors of
the invariant subspace  of $\tb$ in the spectrum of the adjoint
operator $H^\dag$. As we have established in Sec.~\ref{sec:lin},
any eigenvector  of $H^\dag$ from  the invariant subspace of $\tb$
can be represented as a linear combination of $(\tbw^{(1)})^*$ and
$(\tbw^{(2)})^*$.  Requiring the right hand side of  Eq.~(\ref{eq:vep3})  to be
orthogonal to an arbitrary  linear combination of $(\tbw^{(1)})^*$
and   $(\tbw^{(2)})^*$, we arrive at  the following relations:
\begin{equation}
\label{eq:solv}
\frac{\langle F(\tbw(\theta))\tbw(\theta), (\tbw^{(1)})^*\rangle}{\langle\tbw(\theta), (\tbw^{(1)})^*\rangle} = \frac{\langle F(\tbw(\theta))\tbw(\theta), (\tbw^{(2)})^*\rangle}{\langle\tbw(\theta), (\tbw^{(2)})^*\rangle} = B_2.
\end{equation}
In Eq.~(\ref{eq:solv}) the first equality sign is an equation which is to be solved with respect to $\theta$. Once  a root  $\theta$ of the latter equation is found,  then $B_2$ is given from the second equality sign. Notice that despite
the fact that the
vector $\tbw(\theta)$ is complex,
the coefficient  $B_2$ will be real~\cite{ZK}.

Substituting the expression  for $\tbw(\theta)$ into
Eq.~(\ref{eq:solv}), one obtains a rather cumbersome equation,
which, however can be attacked with  a computer algebra program.
After some transformations, Eq.~(\ref{eq:solv}) yields the
following  condition:
\begin{equation}
\label{eq:thetaell}
|e^{2i\theta}| = \frac{|\gamma - \sqrt{-k^2+\gamma^2}|}{{k}}.
\end{equation}
The latter equation has a real root $\theta$  only if  $0<\gamma
\leq k$, i.e. when the expression under the radical is not
positive. This result suggests that there exists a critical value of the gain-loss parameter which we term as the secondary critical point $\ggcr=k$, such that  for sufficiently small
$\gamma$, namely, $0<\gamma \leq \ggcr$,  there exists another family
bifurcating from the eigenvalue $\tb$ of the linear spectrum.
However, this family disappears for $\gamma>\ggcr$, in spite
of the fact that the
$\PT$ symmetry of the underlying linear problem remains unbroken,
i.e. $\ggcr<\gamma_{cr}^{(1)}$. It is important to point that the newly
found  family of solutions does {\it not} correspond to equal
amplitude among the different nodes, and hence pertains to an
elliptically (rather than circularly) polarized family of modes.


Computing the corresponding value of  $B_2$ for  the newly found
elliptically polarized family,  one finds that the coefficient $B_2$ does
depend on $k$ and $\gamma$ (in contrast to the above considered
circularly polarized family, characterized by $B_2=5/3$ for any combination of
$k$ and $\gamma$ which does not violate $\PT$ symmetry).  In
particular, when $\gamma$ approaches $k$, the coefficient $B_2$  of the elliptically polarized family      tends to
$5/3$, which suggests that at $\gamma=k$  the circularly polarized and
elliptically polarized   families  globally merge.

\subsection{Algebraic analysis and numerical results}
\subsubsection{Exact solutions}
Having explored nonlinear modes close to the linear limit, where
the   amplitudes of the modes are small and therefore they can be
analyzed by means of  perturbation theory, let us now consider
nonlinear modes of  arbitrary amplitudes (turning again to the
general case $\alpha\neq0$). Relying on results of the previous
subsection we firstly search for nonlinear modes which have equal
intensities in all four waveguides.  Making the substitution
$w_2=-iw_1^*$ and imposing the condition $w_4=w_1^*$,
$w_3=w_2^*$, which is necessary for a nonlinear mode to be an
eigenstate of $\PT$ thus leading to the circularly polarized light
in each of the coupler arms, system (3)   yields  the
single (complex) algebraic equation
\begin{equation}
    bw_1=k(1-i)w_1^* + \frac{5-\alpha}{3}|w_1|^2w_1-i\gamma w_1.
\end{equation}
Representing $w_1=\rho e^{i\phi}$, we obtain a bi-quadratic equation for $\rho$ yielding two families of the modes bifurcating from the eigenvalues $\tb_\pm$, given by (\ref{b}), of the linear spectrum:
\begin{eqnarray}
\label{eq:rho}
\rho^2_\pm = \frac{3(b  - \tb_{\pm})}{5-\alpha},
\quad
e^{2i\phi_\pm}= \frac{\tb_\pm(1-i) - \gamma(1+i)}{2k} .
\end{eqnarray}
%
Respectively, the nonlinear modes have the following form:
\begin{equation}
\label{eq:symm1}
\bw = \left(\!\!
         \begin{array}{c}
          \rho_\pm e^{i\phi_\pm}\\
          -i\rho_\pm e^{-i\phi_\pm} \\
          i\rho_\pm e^{i\phi_\pm}\\
          \rho_\pm e^{-i\phi_\pm}
         \end{array}
       \!\right)
\end{equation}

Using Eqs.~(\ref{eq:rho})
one can easily obtain continuous families of
nonlinear modes that can be identified as a function of the propagation
constant $b$, for given  $k$ and $\gamma$.

Let us now recall that  Eq.~(\ref{eq:thetaell})  predicts that
for $\alpha=0$ and $\gamma<\ggcr$,
there exist  families of elliptically polarized modes having different  absolute values of the polarization vectors.  Such families were indeed found in our numerics.
%
However, all such modes turned out to be unstable (see
Figs.~\ref{fig:fams}--\ref{a0b2k1} and discussion below).

In the case of zero propagation constant mismatch, i.e. when $\alpha=1$,  one also can find  families which have different amplitudes of the polarization vectors. The explicit expressions for  the families bifurcating from $\tb_\pm$  read
\begin{eqnarray}
\begin{array}{c}
  w_1=w_4^* = \rho_\pm e^{i\phi}, w_2=w_3^* = (-1 \pm \sqrt{2}) \rho_\pm  e^{-i\phi},
\\
    \displaystyle{
    \rho^2_\pm = \frac{b - \tb_\pm}{4 \mp 2\sqrt{2}},\quad
    \phi = \mp \frac{1}{2}\arcsin\frac{\gamma}{\sqrt{2}k}
    }.
\end{array}
    \label{eq:diffamp}
\end{eqnarray}
Remarkably, these modes, which  also describe propagation of elliptically polarized light, are stable in a certain range of the parameters.

\subsubsection{Families of nonlinear modes}

\begin{figure}[tb]
\centerline{
\includegraphics[width=\columnwidth]{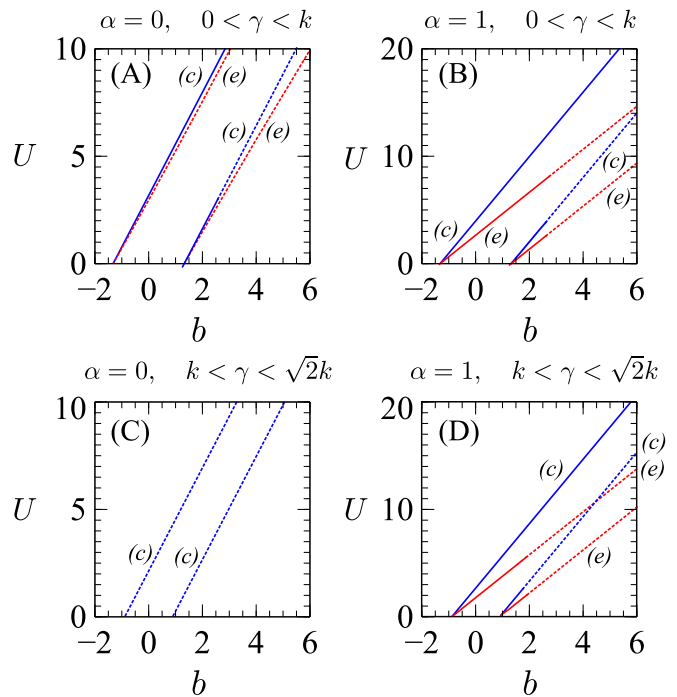}}
\caption{(Color online) Prototypical examples of families of nonlinear modes  in the plane $(b,U)$  for $k=1$ and gain-loss parameters $\gamma$:  $\gamma=0.5$ (the upper panels), $\gamma=1.1$ (the lower panels). Left and right columns correspond to $\alpha=0$ and $\alpha=1$. Stable and unstable modes are shown by by continuous and broken lines, respectively. The families with the circular and elliptical polarization (if any) are marked  with  labels ``\textit{(c)}'' and ``\textit{(e)}'', respectively  (in the color online version families with the circular and elliptical polarization  are  also shown by blue and red lines, respectively).}
\label{fig:fams}
\end{figure}

The results of our analysis of the families of  nonlinear modes  are
summarized in Fig.~\ref{fig:fams}. The upper panels  show  that for $0<\gamma<k$
each eigenvalue of the linearized problem gives rise to two
distinct (circularly and elliptically polarized) families of
nonlinear modes. In the case of $\alpha=0$ the slopes of the dependencies
$U(b)$ are close for the  families of both types, and the
elliptically polarized families are always unstable while
circularly polarized families have both stable and unstable
solutions.  For $\alpha=1$ one can find stable solutions both for the
families with circular and for those with elliptical polarization.

For $k<\gamma<\sqrt{2}k$, the case $\alpha=0$ does not allow for   elliptically polarized families [see Fig.~\ref{fig:fams}~C], a feature which is in accordance with the perturbation approach developed above. In this case one can only find circularly polarized modes, which are unstable. On the other hand, for $\alpha=1$, stable and unstable modes of both types can be found [see Fig.~\ref{fig:fams}~D].


%
%

Summarizing at this point, we have identified 4 sets of
solutions, two circularly polarized with equal amplitude at the nodes,
and two elliptically polarized with unequal such amplitudes.
These all degenerate into the two distinct eigenvalues
$\tb_\pm$, given by (\ref{b}),  of the linear problem.
The circularly polarized solutions  are more robust, while the elliptically polarized ones
are always unstable for $\alpha=0$ and stable only for
small enough amplitudes for $\alpha=1$. Among the   circularly polarized
ones, for $0<\gamma<\sqrt{2}k$ the more fundamental state (stemming from the negative eigenvalue
at the linear limit) is always the stable ground state   of
the system in continuations over the parameter $b$, while the
excited state is only stable for small enough amplitudes.

\subsubsection{Continuation over $\gamma$}

\begin{figure}[htp]
\scalebox{1.04}{\includegraphics[trim=1.8cm 1.5cm 2.5cm 0.9cm, clip=true,width=\columnwidth]{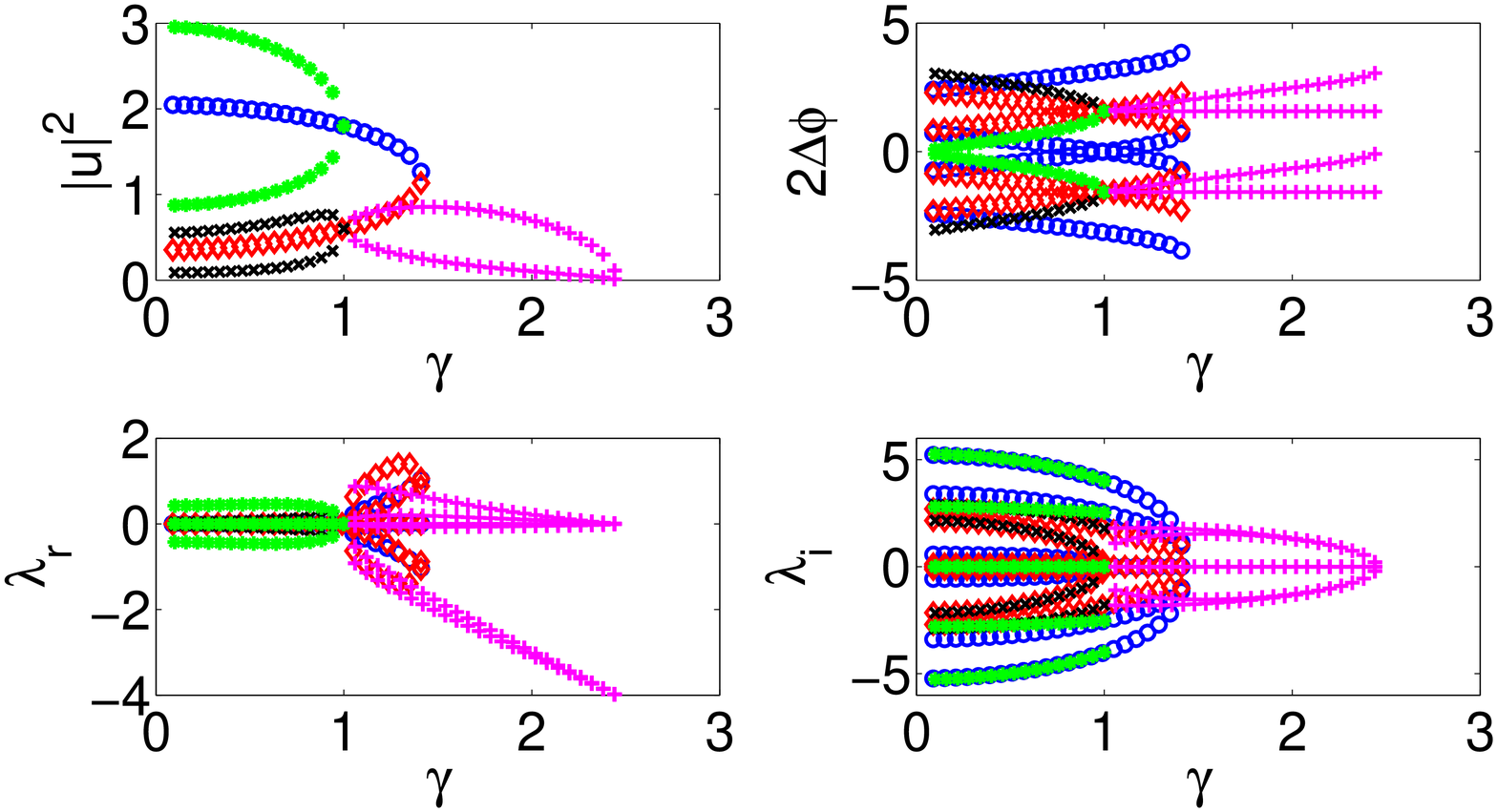}}
\caption{(Color on-line) The four panels denote the solution amplitude
(top left), phase differences between adjacent nodes
(top right),  real and imaginary
parts (second row) of eigenvalues for $\alpha=0$, $b=2$, and $k=1$.
The blue circles branch corresponds to the ``+'' sign in Eqs.~(\ref{eq:rho}),
while the red
diamonds branch corresponds to the ``-'' sign (the symmetric intensity or
circularly polarized
branches). 
The green stars and black squares crosses are those solutions with distinct
absolute values of the
polarization vectors (the asymmetric or elliptically polarized
branches). In the top left panels, they collide and disappear
in two subcritical pitchfork bifurcations with the blue circles
and red diamond branches, respectively.
The magenta pluses branch in the
panels represents the ghost state solutions, which bifurcate from
the red diamonds at $\gamma=\ggcr=1$ and terminate at $\gamma=2.44$.
}
\label{a0b2k1}
\end{figure}

An alternative and perhaps even more telling way to illustrate the
above  features stems from fixing some value of $b$, starting
from the Hamiltonian limit of $\gamma=0$ and subsequently
identifying branches of the nonlinear modes by means of changing
$\gamma$, as shown in  Fig.~\ref{a0b2k1}. It is important to note
that this alternative viewpoint affords us the ability to
visualize bifurcations which we now explore.

\begin{figure}[htp]
\scalebox{1.04}{\includegraphics[trim=1.8cm 1.5cm 2.5cm 0.9cm, clip=true,width=\columnwidth]{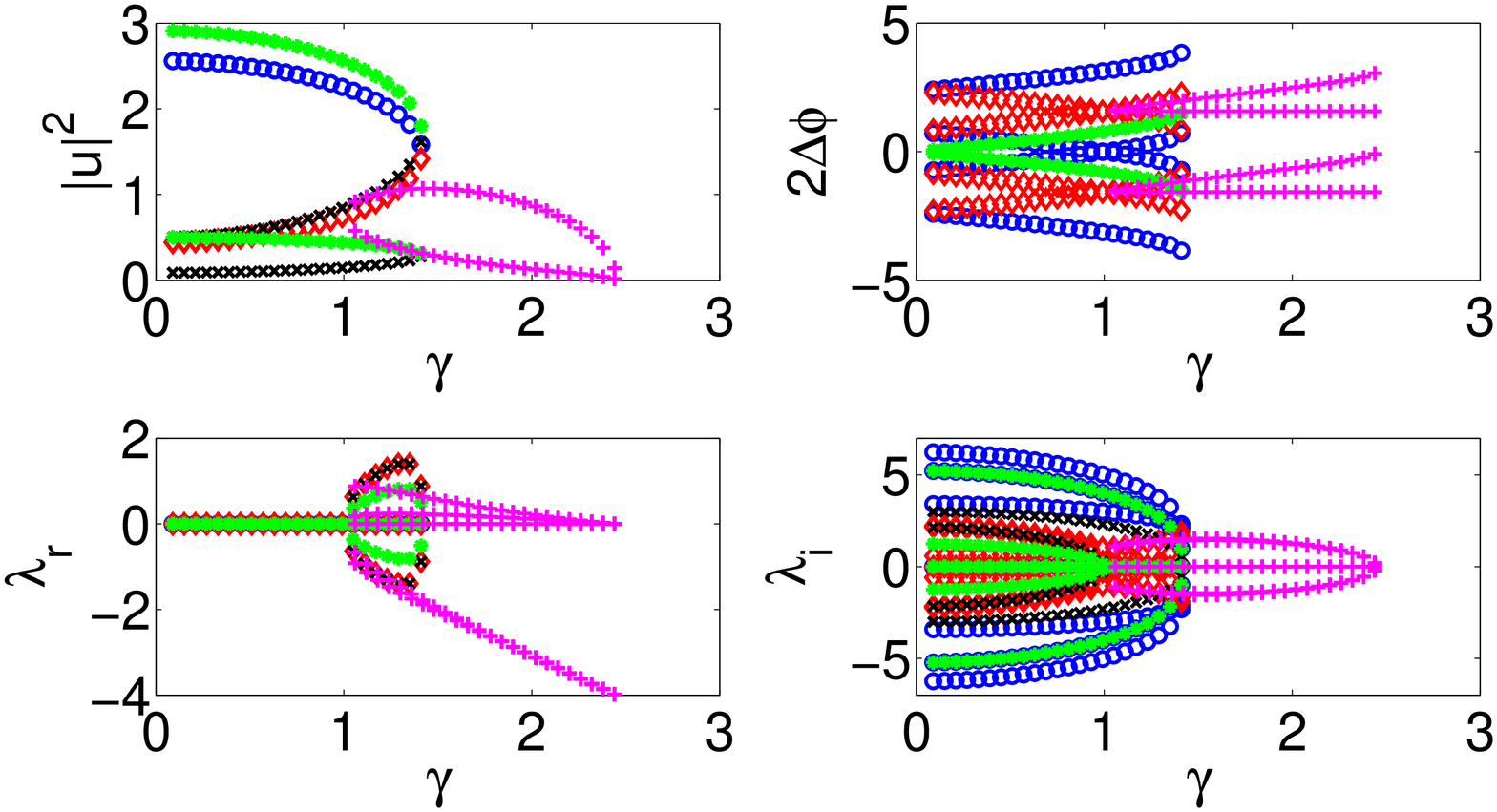}}
\caption{The four panels show the same diagnostics as in the
previous figure
but now for $\alpha=1$, $b=2$, and $k=1$.
}
\label{a1b2k1}
\end{figure}

\begin{figure}[htp]
\includegraphics[width=0.49\columnwidth]{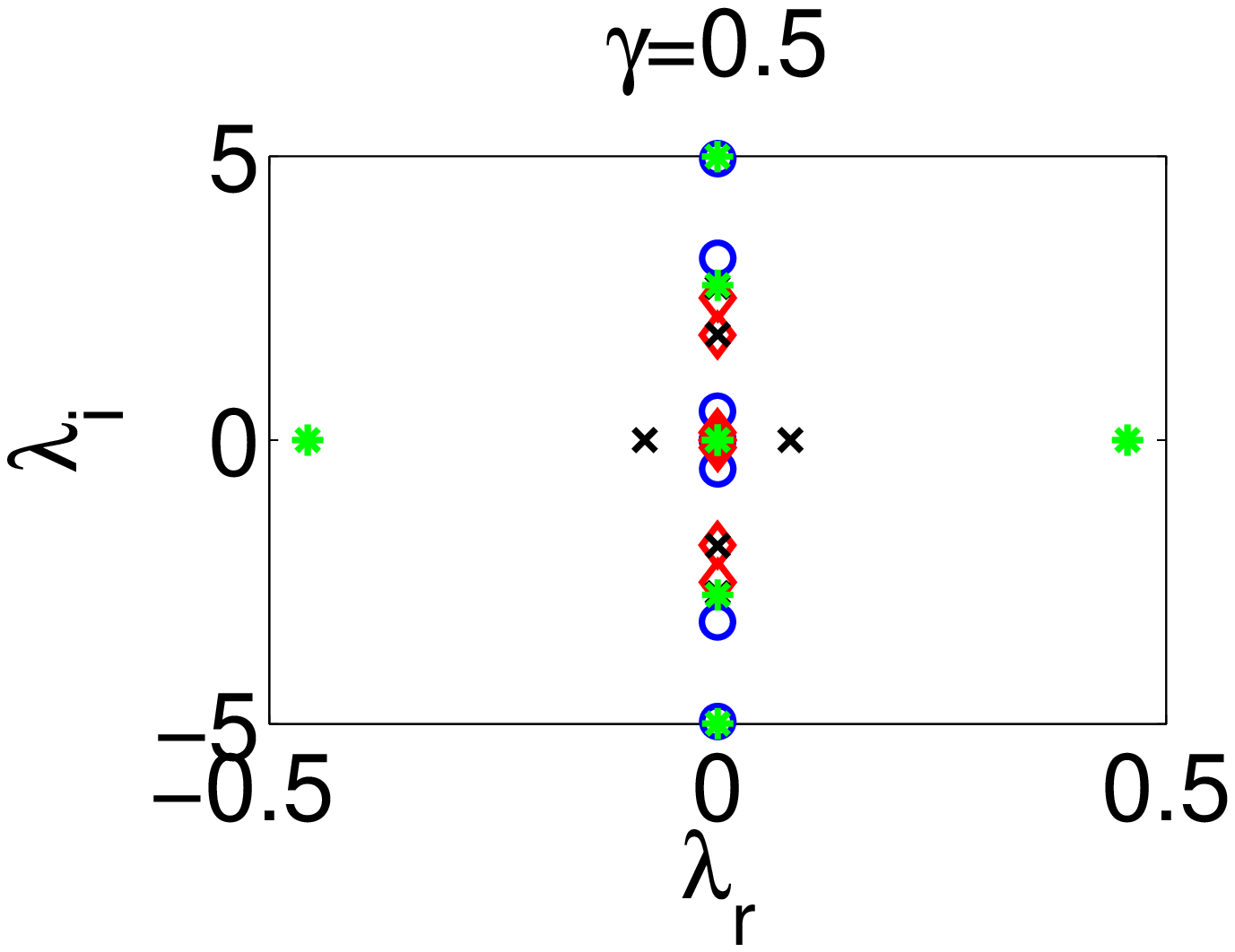}
\includegraphics[width=0.49\columnwidth]{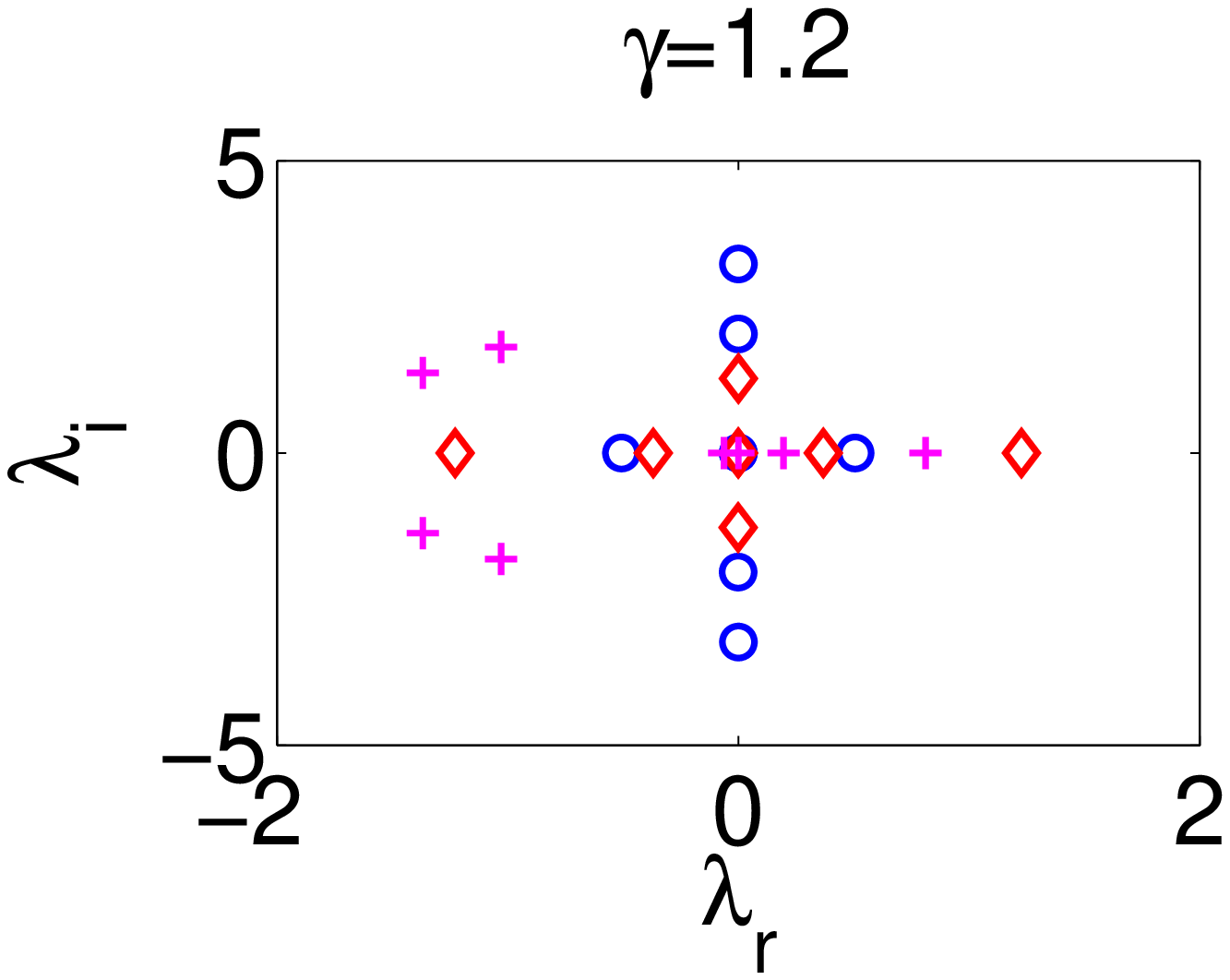}
\includegraphics[width=0.49\columnwidth]{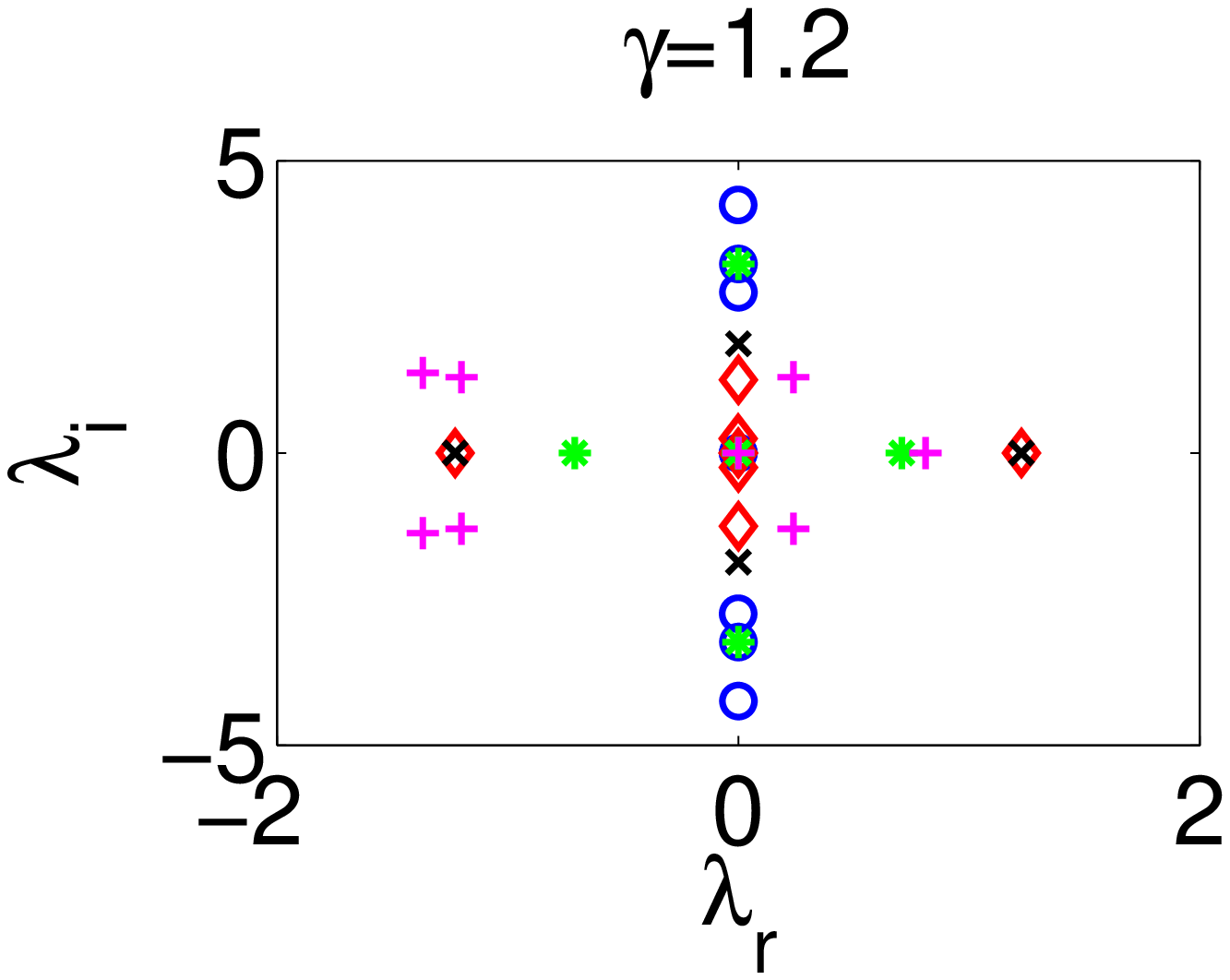}
\caption{Stability plots. The top two panels are for $\alpha=0$, and the bottom one is when $\alpha=1$.
In the case $\alpha=0$, at $\gamma=\ggcr=1$,  one pair of blue circles and two pairs of red diamonds collide at 0 so that
one pair of real eigenvalues arises in blue circles branch whereas two
pairs of real eigenvalues arise in the red diamonds branch.
The asymmetric branches only exist (and are unstable) for the smaller
value of $\gamma=0.5$, for $\alpha=0$. For the same parameters
($b=2$, $k=1$),
in the
case of $\alpha=1$, the excited symmetric and both asymmetric branches
are unstable for $\gamma=1.2$.}
\label{a0b2k1_s}
\end{figure}

The relevant results for parametric continuations over $\gamma$
are given in Figs.~\ref{a0b2k1}-\ref{a1b2k1}; typical examples of
the corresponding linearization spectra for different values of
$\gamma$ can be found in Fig.~\ref{a0b2k1_s}. Here, it can be seen
that a lower amplitude and a higher-amplitude intensity-symmetric (i.e., equal
amplitude) branch
exist, for fixed $b$, from the Hamiltonian limit of $\gamma=0$ and
all the way up to the linear $\PT$-phase transition point $\gamma_{cr}^{(1)}=\sqrt{2} k$. At that point,
the two equal amplitude branches collide and disappear in a
saddle-center bifurcation which can be thought of as a nonlinear
analog of the linear $\PT$-phase transition~\cite{djf}. An
additional very interesting feature arises precisely at the point
$\gamma_{cr}^{(2)}$ [see (\ref{eq:thetaell}) and the related discussion],  where it can be seen that both branches of equal
amplitude between the sites become dynamically unstable for
$\alpha=0$.
In fact, it is seen that for the larger amplitude branch (associated
with the blue circles), one pair of unstable eigenvalues arises, while
for the smaller amplitude (red diamond) branch, two such pairs accompany
the symmetry breaking bifurcation
occurring at this critical point. A closer inspection reveals
that the symmetric branch (blue circles) is destabilized through
a subcritical pitchfork bifurcation with its ``corresponding'' asymmetric
state (i.e., the one degenerate with it in the linear limit).
In the case of the lower amplitude (excited) state for the same
$b$, the situation appears to be more
complex. In particular, there exists once again a subcritical pitchfork
with the corresponding asymmetric branch, yet this would justify
one pair of unstable eigenvalues and we observe two. This is because
at the {\it same} point, there also exists a supercritical pitchfork,
which gives rise to the so-called ghost states, denoted by magenta
plus symbols. These states are analogous to the ones to
analyzed in~\cite{R46a,R46b,R46}, but remarkably are not
stationary states of the original problem, yet they are pertinent
to its dynamical (instability) evolution and for this reason they
will be examined in further detail separately in the dynamics section below.

In the case of $\alpha=1$, only one pair of unstable eigenvalues
emerges for the lower amplitude branch at the secondary
critical point of $\gamma_{cr}^{(2)}$ (while the larger amplitude
branch remains stable throughout the continuation in $\gamma$).
Hence, in this case, once again a saddle-center bifurcation
will mark the nonlinear $\PT$-phase transition, yet the number
of unstable eigendirections of each symmetric
branch (fundamental and excited) is decreased by one (0 and 1
real pairs instead of 1 and 2, respectively, for $\alpha=0$).
In this case, in fact, both asymmetric branches persist up
to the linear $\PT$-phase transition (rather than terminate
in a subcritical pitchfork as above), and collide and disappear
with each other. Interestingly all 3 branches (the lower amplitude,
excited symmetric one and the two asymmetric ones) become unstable
at the secondary critical point $\ggcr=k$,
which again points to the existence
of corresponding ghost states.
For the lower amplitude symmetric branch, the bifurcating ghost states
are again identified by the magenta
plus symbols
in Fig.~\ref{a1b2k1}.

\section{Dynamics of the polarization}
\label{sec:stab}

\begin{figure}[htp]
\includegraphics[width=\columnwidth]{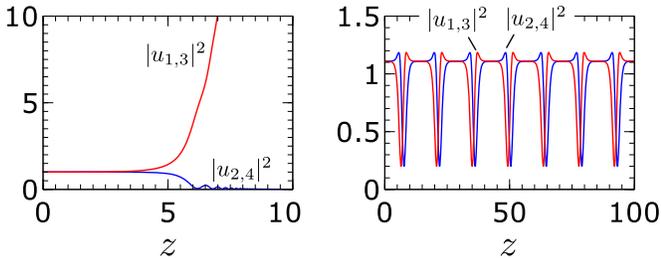}%
\caption{(Color online)  Dynamics of an unstable circularly polarized mode at $b=3$, $\gamma=0.5$ and
for $\alpha=0$ for two different  small initial perturbations.
The result of the evolution of the left panel involves
growth at the gain sites ($u_{1,3}$, red curves in the color online version) and decay at the lossy sites ($u_{2,4}$, blue curves in the color online version).  Notice that intensities among the two gain sites and
among the two lossy sites are approximately equal ($|u_1|^2\approx |u_3|^2$ and $|u_2|^2\approx |u_4|^2$) and are not distinguishable in the scale of the plots.
In the right panel only the initial stage of the found persistent periodic
dynamics is shown; the simulations were performed up to $z=2000$.
} \label{fig:dyn1}
\end{figure}

\begin{figure}[htp]
\includegraphics[width=0.5\columnwidth]{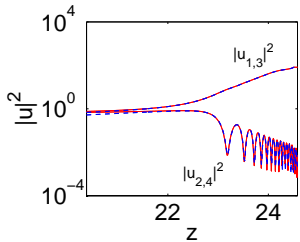}
\caption{
The dynamical semi-log plot of the ghost state
with $\alpha=0$, $b=2$, and $k=1$ for $\gamma=1.02$. The red solid
line and blue dashed line correspond to the red diamond and
magenta plus branches in Fig.~\ref{a0b2k1}, respectively. The
blue dashed line is plotted with a shift in time, i.e. delay by
$z=19.61$. The time axis in the plot is the actual time of the red
solid line. Their dynamical behaviors are essentially identical
(and
can not be distinguished on the scale of the plot), showcasing how
the unstable dynamics of the red diamond branch is similar to 
that of the bifurcating ghost.}
\label{a0b2k1_ghostb}
\end{figure}

To examine the potentially symmetry breaking (and more generally
instability driven) nature of the dynamical evolution
past the critical points identified above, we have also performed direct
numerical simulations which are illustrated in Fig.~\ref{fig:dyn1};
see also Fig.~\ref{a0b2k1_ghostb}.
Here, it can be seen that although the relevant parameters are
below the critical point for the linear $\PT$-phase transition
$\gamma_{cr}^{(1)}=\sqrt{2} k$, nevertheless, symmetry breaking phenomena
are observed due to the dynamical instability of the relevant
states (the ones denoted by dashed  lines in Fig.~\ref{fig:fams}).
This dynamics may, in principle, be associated with the so-called
ghost states of complex propagation constant that have
recently been proposed as relevant for the dynamical evolution
in~\cite{R46}.
To substantiate this claim, we note that it is observed in the
left panel of Fig.~\ref{fig:dyn1} that the relative phase of the
two gain sites that lock into an equal growing amplitude, is
$\pi/2$, as is those of the decaying amplitude lossy sites. In
light of this, we seek ghost states with precisely this phase
difference and are able to explicitly identify them via the ansatz
$w_3=iw_1,\ w_4=iw_2$, setting $w_j=c_je^{i\phi_j}$ for $j=1,2$. For these branches, the
propagation constant is complex.
This highlights the potential growth or decay of such
states. Importantly also, note that these states are ``ghosts''
because they may be solving the stationary problem of
Eqs.~(\ref{static}), but the U$(1)$ invariance of the original
model does {\it not} permit them to be a solution of the dynamical
Eqs.~(\ref{dynam}).

The algebraic conditions that this
family of solutions satisfies are
\begin{align}
&\sin\phi_b=\frac{(c_2^2-c_1^2)\gamma}{(c_1^2+c_2^2)B} \\
&\cos\phi_b=\frac{(5-\alpha)(c_1^2+c_2^2)}{3B} \\
&\sin(\phi_2-\phi_1) \nonumber \\
=&\frac{(3\gamma-B(\sin\phi_b+\cos\phi_b)+(5-\alpha)c_2^2)c_2}{6kc_1} \\
=&\frac{(3\gamma+B(\sin\phi_b-\cos\phi_b)+(5-\alpha)c_1^2)c_1}{6kc_2} \\
&\cos(\phi_2-\phi_1) \nonumber \\
=&\frac{(3\gamma-B(\sin\phi_b-\cos\phi_b)-(5-\alpha)c_2^2)c_2}{6kc_1} \\
=&\frac{(3\gamma+B(\sin\phi_b+\cos\phi_b)-(5-\alpha)c_1^2)c_1}{6kc_2}.
\end{align}
Notice that the imaginary part of the propagation constant $B \sin(\phi_b)$
is proportional to the difference $c_1^2-c_2^2$. Hence, prior to the
symmetry breaking, the relevant solutions bear a real propagation
constant. Past the bifurcation point
one (unstable) branch has $c_2^2>c_1^2$, while the stable branch
has $c_1^2>c_2^2$.
The relevant ghost state branches and their bifurcation from the
equal amplitude ones
are explored in Fig.~\ref{a0b2k1}-\ref{a1b2k1}. 
Given that these are only ghost solutions of the original dynamical
problem, the interpretation of their linearization spectrum
(shown for completeness in Fig.~\ref{a0b2k1_s}) is still an open problem.

These ghost states appear, in fact, to exhibit very similar 
evolution dynamics to regular unstable states.
To illustrate this, we
observed the particular behavior of the unstable modes
and how it depends on the form of the initial perturbation.
A typical example in which the gain sites lead
to growth and the lossy sites to decay is shown in the
left panel of Fig.~\ref{fig:dyn1}. It is interesting that 
the evolution appears to be very proximal to that of the 
ghost states identified above.
This is clearly showcased
in Fig.~\ref{a0b2k1_ghostb}, through the comparison of the
growth pattern in the relevant sites (and the decay pattern in the
lossy sites) with the exact, shifted in the propagation distance to fit the onset of
growth, ghost state solution for the same parameters. 
On the other hand, in the right panel of Fig.~\ref{fig:dyn1}
a different scenario of
evolution is illustrated. Instead of the gain
nodes growing and the lossy ones decaying, a breathing oscillation
settles between the two pairs. These two scenarios, illustrated
in Fig.~\ref{fig:dyn1},  are the prototypical instability evolution
ones that we have obtained in this system.



\section{Conclusion}
\label{sec:Concl}

In conclusion, in the present work, we have proposed a novel,
physically realistic variant of a $\PT$ symmetric dimer
where the effect of birefringence has been taken into consideration.
The existence of polarization of the electric field within the
coupler yields two complex dynamical equations for each of the fibers,
providing a physical realization of a plaquette model with
{\it both linear and nonlinear} coupling between the elements.
The stationary states of the model were identified and both
linear and nonlinear $\PT$-phase transitions were obtained.
The degenerate nature of the linear limit complicated the
problem in comparison to other ones studied earlier in this context.
Furthermore, the emergence of symmetry breaking phenomena
and associated (subcritical or supercritical) pitchfork bifurcations,
as well as their dynamical implications in leading to indefinite growth
and decay (of the corresponding waveguide amplitudes) were elucidated.
A connection was also given to ghost states.

From the physical point of view, we emphasize that the
use of $\PT$ symmetry significantly changes the stability of the modes with
different polarizations.
For example, in the case of large propagation constant mismatches,
we saw that the symmetric (circularly polarized) states
become destabilized in the presence of gain/loss and may even cease
to exist past a certain critical point of the relevant parameter.
Instead of them, the dynamics may lead to a breathing exchange of
``mass'' between the gain and the lossy waveguides, or most commonly
an indefinite growth of the former at the expense of the latter. 
This is in line with what is observed also in the evolution of the 
emerging ghost states of the system. The instabilities and
associated dynamics should be observable in suitable generalizations
of existing experiments such as~\cite{kip}.
Such  properties are relevant also to possibilities of solitonic 
(waveguide-array) generalizations of the coupler system as well as towards a possible use
of $\PT$ symmetric coupler for measurement techniques based on the use of
several nonlinear modes.

From the mathematical point of view, we believe that these studies may pave the way for considering
multi-component, as well as multi-dimensional
(generalizing the plaquettes considered here or those
of~\cite{guenther}) lattice models of $\PT$-symmetric form.
In generalizing to multi-plaquette
configurations, it would be especially interesting to examine
which of the symmetry-breaking and nonlinear
$\PT$-phase transition phenomena examined herein are preserved
and what new phenomena may arise as additional degrees of freedom
are added. Such studies will be deferred to future publications.

\acknowledgments

VVK and DAZ acknowledge support of the FCT
grants PTDC/FIS/112624/2009, PEst-OE/FIS/UI0618/2011, and SFRH/BPD/64835/2009.
PGK gratefully acknowledges support from the Alexander von Humboldt
Foundation, and
the Alexander S. Onassis Public Benefit Foundation, as well as from
the US-NSF (grants DMS-0806762, CMMI-1000337) and from the US-AFOSR.

\end{document}